\documentclass[12pt,a4paper]{iopart}
\usepackage{iopams}
\newcommand{\Threej}[6]{\pmatrix{
  {#1} & {#2} & {#3}\cr
  {#4} & {#5} & {#6}\cr}}
\newcommand{\Sixj}[6]{\left\{\matrix{
  {#1} & {#2} & {#3}\cr
  {#4} & {#5} & {#6}\cr}\right\}}
\begin{document}

\input{epsf.sty}
\jl{2}
\title[Collisional loss of atoms from harmonic traps]
{Ultracold atomic collisions in tight harmonic traps:
    Perturbation theory, ionization losses and application to
    metastable helium atoms}

\author{T J Beams\dag, G Peach\ddag\ and I B Whittingham\dag}

\address{\dag\ School of Mathematical and Physical Sciences,
James Cook University, Townsville, Australia, 4811}

\address{\ddag\ Department of Physics and Astronomy, University College
London, Gower Street, London, WC1E 6BT, UK}

\begin{abstract}
Collisions between tightly confined atoms can lead to ionization and hence to
loss of atoms from the trap.  We develop second-order perturbation theory for a
tensorial perturbation of a spherically symmetric system and the theory is then
applied to processes mediated by the spin-dipole interaction.  Redistribution
and loss mechanisms are studied for the case of spin-polarized metastable
helium atoms and results obtained for the five lowest $s$ states in the 
trap and
trapping frequencies ranging from 1 kHz to 10 MHz.
\end{abstract}


\maketitle

\section{Introduction}

There is significant interest in the study and control of quantum processes
involving trapped ultracold neutral atoms where the trapping environments are
so tight that the effect of the trapping fields upon the colliding atoms cannot
be ignored or approximated as constant background fields.  Trapping in
three-dimensional optical lattices, with typical trapping frequencies of
$10^{4}$ to $10^{6}$ Hz, forms the basis of such studies as quantum phase
transitions of ${}^{87}$Rb atoms \cite{Greiner2002}, storage of metastable
argon atoms \cite{Muller1997}, implementation of quantum logic gates and
formation of highly entangled quantum states \cite{Brennen1999,Jaksch1999}.
Theoretical investigations have focussed mainly on tightly confined alkali
systems and have been based either upon direct numerical integration of the
radial Schr\"odinger equation for the relative motion of the colliding atoms
using the best available full interatomic potentials \cite{Tiesinga2000} or
upon a regularized delta-function pseudopotential and an energy-dependent
effective scattering length \cite{Bolda2002,Blume2002}.

An understanding of collision processes in trapped ultracold metastable $2^3$S
helium (denoted by He$^*$) is necessary to obtain Bose-Einstein condensation of
this species \cite{Santos2001,Robert2001} and to investigate these novel
excited-state condensates \cite{Sirjean2002,Seidelin2002}.  Although current
experiments on He$^*$ only use trapping frequencies of the order of 10$^2$ to
10$^3$ Hz, it is of interest to investigate the effects of much tighter
trapping on the allowed quantized trap states, as a possible tool to manipulate
the confined atoms, and to enhance trap loss through ionization processes at
small interatomic separations as a means of studying these processes.

We have recently analyzed a system of two colliding ultracold atoms under
strong harmonic confinement in a spherically symmetric trap from the viewpoints
of quantum defect theory and of elastic scattering in the interatomic
potential.  We have developed methods for determining the energies of the
quantized states produced by the presence of the trap and the theory was
applied to collisions between spin-polarized He$^*$ atoms, see Peach \etal
\cite{CCP62002,Peach2004}.  The energies were determined for a wide range of
trapping frequencies for $s$- and $d$- wave collisions using two totally
independent methods to integrate the radial Sch\"odinger equation.  Excellent
agreement was obtained between the two methods, one based on the use of quantum
defect theory and the second on the use of a discrete variable representation.

These calculations ignored loss processes, but inelastic collisions may cause
transitions to states from which there is a high probability of Penning and
associative ionization.  A study of such loss processes is the subject of this
paper which is organized as follows.  In section 2, the theory of collisions in
an isotropic trap is briefly reviewed, and in section 3, second-order
perturbation theory is introduced for a general form of the perturbation and
for trap states of any angular momentum.  In section 4, the theory is applied
to perturbation by the spin-dipole interaction and in section 5 the numerical
methods are described.  Finally in section 6 the theory is applied to the case
of spin-polarized He$^*$ atoms.  It is found that only the $s$ states are
significantly perturbed and shifts and lifetimes are presented for the five
lowest $s$ states in the trap for trapping frequencies ranging from 1 kHz to
10 MHz.

\section{Collisions between two atoms in an isotropic harmonic trap}

Consider two atoms with masses $M_1$ and $M_2$, spin quantum numbers $S_1$ and
$S_2$ and position vectors $\bi{r}_1$ and $\bi{r}_2$ relative to the
centre of the trap.  The interatomic separation is given by
$r=|\bi{r}|=|\bi{r}_{1}-\bi{r}_{2}|$ and, for the case of an
atom-atom potential that is only a function of $r$ combined with a potential
for an isotropic harmonic trap of angular frequency $\omega$, the Hamiltonian
is separable into two parts $H_{\rm{cm}}$ and $H_0$ describing the
centre-of-mass and relative motions of the two atoms.  Here we will use and
extend the notation developed in \cite{Peach2004}.  If the total spin quantum
number is $S$ and the adiabatic potential for the molecular state
${}^{2S+1}\Lambda$ is denoted by $V_{\Lambda S}(r)$, the equation for the
relative motion is
\begin{equation}
\label{eigen}
    H_0|\psi(\bi{r})\rangle = E |\psi(\bi{r})\rangle \,,
\end{equation}
where $H_0$ is defined by
\begin{equation}
\label{hamil0}
    H_0 \equiv \left[ -\frac{\hbar^2}{2M} \nabla_{r}^{2} +
    \frac{1}{2}M \omega^{2} r^{2} \Delta_{\rm{trap}}
    + V_{\Lambda S}(r) \right] \,.
\end{equation}
In (\ref{eigen}) and (\ref{hamil0}) $E$ is the energy eigenvalue, the 
reduced mass $M =
M_{1}M_{2}/(M_{1}+M_{2})$ and $\Delta_{\rm{trap}}$ = 1 or 0 according to
whether the harmonic potential is turned on or off.  If the angular momentum
quantum number for the relative motion is $l$, the eigenvector
$|\psi(\bi{r})\rangle$ is given by
\begin{equation}
\label{psi}
    |\psi(\bi{r})\rangle = \frac{1}{r} F_{kl\Lambda S}(r)
    |l m\rangle |S_1 S_2 S M_S\rangle \,,
\end{equation}
where the magnetic quantum numbers $m$ and $M_S$ refer to projections
of the angular and spin momenta onto the molecular axis. The radial
function $F_{kl\Lambda S}(r)$ satisfies the equation
\begin{equation}
\label{Fkl}
    \left[\frac{\rmd^2}{\rmd r^2} - \frac{l(l+1)}{r^2} - \frac{r^2}{\xi^4}
    \Delta_{\rm{trap}} - \frac{2 M}{\hbar^2} V_{\Lambda S}(r) + k^2\right]
    F_{kl\Lambda S}(r) = 0 \,,
\end{equation}
where we have introduced the quantities
\begin{equation}
\label{k2xi}
    k^2 = \frac{2ME}{\hbar^2}\,; \qquad \xi^2 = \frac{\hbar}{M\omega} \,.
\end{equation}
   It has also been shown in \cite{Peach2004}
that for the discrete states with $E > 0$ produced by the presence of the
trap, hereafter referred to as trap states, it is natural to introduce an
effective quantum number (or scaled energy) $n^*$, where
\begin{equation}
\label{nstar}
    n^* = \frac{E}{2\hbar\omega} = n'_r + \frac{l}{2} +
    \frac{3}{4} -\mu'\,; \qquad n'_r = 0,1,2, \ldots \,,
\end{equation}
and $\mu'$ is a quantum defect that varies slowly as $n'_r$ increases.
By introducing the dimensionless variable $\rho = r/\xi$, equation
(\ref{Fkl}) can also be written in the form
\begin{equation}
\label{Fklrho}
    \left[\frac{\rmd^2}{\rmd \rho^2} - \frac{l(l+1)}{\rho^2}
    - \rho^2 \Delta_{\rm{trap}}
    - \frac{2 V_{\Lambda S}(\rho)}{\hbar\omega} + 4n^*\right]
    F_{kl\Lambda S}(\rho) = 0 \,.
\end{equation}

\section{Perturbation theory}

The Hamiltonian for the perturbed system is given by
\begin{equation}
\label{htotal}
    H = H_0 + H_{\rm{p}} \,,
\end{equation}
where $H_{\rm{p}}$ is the perturbing potential and it is assumed that
the eigenstates of $H_0$ are known, i.e.
\begin{equation}
\label{H0eigen}
     H_0 |j \rangle = E_j |j \rangle; \qquad j = 0,1,2,\ldots \,.
\end{equation}
Then the change in energy of the initial state $i$, correct to second order,
is given by
\begin{equation}
\label{DelE}
    \Delta E = \Delta E_1 + \Delta E_2 \,,
\end{equation}
where
\begin{equation}
\label{DelE1}
    \Delta E_1 = \langle i|H_{\rm{p}}|i\rangle
\end{equation}
and
\begin{equation}
\label{DelE2}
    \Delta E_2 = - \sum_{j\neq i} \frac{\langle i|H_{\rm{p}}^\dagger|j\rangle
    \langle j|H_{\rm{p}}|i\rangle}{(E_j-E_i)} \,.
\end{equation}
The calculation of $\Delta E_1$ is straightforward and $\Delta E_2$ can be
evaluated as follows.  Using the method of Dalgarno and Lewis
\cite{Dalgarno1955}, we introduce the operator $\hat{F}$ which satisfies the
inhomogeneous equation
\begin{equation}
\label{Fdef}
    \left[\hat{F} H_0 - H_0 \hat{F}\right]|i\rangle = H_{\rm{p}}|i\rangle \,,
\end{equation}
so that (\ref{DelE2}) becomes
\begin{equation}
\label{DelEF}
    \Delta E_2 = \langle i|H_{\rm{p}}^\dagger\; \hat{F}|i\rangle -
    \langle i|H_{\rm{p}}^\dagger|i\rangle \langle i|\hat{F}|i\rangle \,.
\end{equation}

This result is only useful if we can determine the operator $\hat{F}$.
We consider the perturbation $H_{\rm p}$ given by
\begin{equation}
\label{Hp}
    H_{\rm p} = \mathbf{T}(\lambda)\mathbf{\cdot U}(\lambda)\, 
V_{\rm{p}}(r) \,,
\end{equation}
where $\mathbf{T}(\lambda)$ and $\mathbf{U}(\lambda)$ are tensor operators of
order $\lambda$ and $V_{\rm{p}}(r)$ contains the
radial dependence of $H_{\rm p}$.
We set $|i \rangle \equiv |\psi(\bi{r}) \rangle$ and make the following
expansion
\begin{equation}
\label{Fexpan}
\fl
    \hat{F}|i \rangle \equiv
    \hat{F}\,\frac{1}{r}\,F_{kl\Lambda S}(r) |l m \rangle | S M_S \rangle
    = \sum_{l'm'S'M'_S} \frac{1}{r} f_{l'm'S'M'_S}(r) |l'm'\rangle
    |S'M'_S \rangle \,,
\end{equation}
where we have suppressed the spin quantum numbers $S_1$ and $S_2$ in
(\ref{psi}).  Then, on using (\ref{eigen})--(\ref{Fkl}), (\ref{Fdef}),
(\ref{Hp}) and (\ref{Fexpan}), we obtain
\begin{eqnarray}
\label{FH0}
    (\hat{F}H_0 - H_0\hat{F})|i \rangle = \frac{\hbar^2}{2M} \sum_{l'm'S'M'_S}
    \nonumber \\ \times
    \left[\frac{\rmd^2}{\rmd r^2} - \frac{l'(l'+1)}{r^2} -
    \frac{r^2}{\xi^4} \Delta_{\rm{trap}} - \frac{2M}{\hbar^2} V_{\Lambda'S'}
    + k^2\right] f_{l'm'S'M'_S} |l'm' \rangle |S'M'_S \rangle
    \nonumber \\
    = \mathbf{T}(\lambda)\mathbf{\cdot} \mathbf{U}(\lambda)
    \,V_{\rm{p}}\,F_{kl\Lambda S} |lm \rangle |SM_S \rangle \,.
\end{eqnarray}
Now if we define $G_{kl'\Lambda'S'}(r)$ by the relation
\begin{equation}
\label{Gkldef}
    f_{l'm'S'M'_S}(r) = \langle S'M'_S| \mathbf{T}(\lambda)|S M_S \rangle
    \mathbf{\cdot}\langle l'm'| \mathbf{U}(\lambda) |lm \rangle
    \; G_{kl'\Lambda'S'}(r) \,,
\end{equation}
then $G_{kl'\Lambda'S'}(r)$ satisfies the inhomogeneous radial equation
\begin{equation}
\label{Gkl}
\fl
    \left[\frac{\rmd^2}{\rmd r^2} - \frac{l'(l'+1)}{r^2}
    - \frac{r^2}{\xi^4} \Delta_{\rm{trap}}
    - \frac{2M}{\hbar^2} V_{\Lambda'S'}(r) + k^2 \right]
    G_{kl'\Lambda'S'}(r) = \frac{2M}{\hbar^2}
    V_{\rm{p}}(r)\,F_{kl\Lambda S}(r) \,,
\end{equation}
c.f. (\ref{Fkl}).  On introducing the scaled variable $\rho$, (\ref{Gkl})
becomes
\begin{eqnarray}
\label{Gklrho}
\fl
    \left[\frac{\rmd^2}{\rmd \rho^2} - \frac{l'(l'+1)}{\rho^2}
    - \rho^2 \Delta_{\rm{trap}}
    - \frac{2 V_{\Lambda'S'}(\rho)}{\hbar\omega} + 4n^* \right]
    G_{kl'\Lambda'S'}(\rho) =
    \frac{2 V_{\rm{p}}(\rho)}{\hbar\omega}\,F_{kl\Lambda S}(\rho)
    \nonumber \\ \lo
    \equiv W(\rho)\,F_{kl\Lambda S}(\rho) \,,
\end{eqnarray}
c.f. (\ref{Fklrho}).  If $\Delta_{\rm{trap}} = 0$ in (\ref{Fkl}),
then in the outer region where $V_{\Lambda S}(r)$ is very small,
the regular and irregular solutions for $F_{kl\Lambda S}(r)$ are
given by
\begin{eqnarray}
\label{FRFI}
    F^{\rm{R}}_l(r) \simeq N\,(kr)[\cos \delta_{l}\;
    j_{l}(kr) - \sin \delta_{l} \;n_{l}(kr)] \,; \nonumber \\
    F^{\rm{I}}_l(r) \simeq - N\,(kr)[\sin \delta_{l}\;
    j_{l}(kr) + \cos \delta_{l} \;n_{l}(kr)] \,,
\end{eqnarray}
where $j_{l}(kr)$ and $n_{l}(kr)$ are spherical Bessel functions, see
\cite{Abram1965}, $N$ is a normalization constant and $\delta_{l}
\equiv \delta_{l}(k)$ is the $l$-wave phase shift for elastic
scattering. The wave function $F_{kl\Lambda S}(r)$ is matched to a
normalized bound-state wave function of the same energy by choosing
\begin{equation}
\label{normal}
    N = 2\, \left[\pi\xi^2 k \left(1+\frac{\rmd \mu'}{\rmd n^*}
          \right) \right]^{-\frac{1}{2}} ,
\end{equation}
c.f. (\ref{nstar}), where $\mu' \equiv \mu'(n^*)$ is treated as a
continuous function.  Also, as $r \rightarrow \infty$, it can be shown
that the contribution to the solution of (\ref{Gkl}) from the particular
integral is given by
\begin{equation}
\label{partint}
    G_{kl'\Lambda'S'} = \Re\left\{[F^{\rm{R}}_l(r) +
    \rmi\,F^{\rm{I}}_l(r)]\; \frac{1}{r^2} R(x)\right\} \,;
    \qquad x \equiv \frac{1}{r} \,,
\end{equation}
where $R(x)$ is a slowly varying complex function of $x$ satisfying
the conditions
\begin{equation}
\label{condns}
    R(0) = \frac{M}{2k}\;\rmi\,;\qquad \left.
    \frac{\rmd R}{\rmd x}\right|_{x=0} = 0 \,.
\end{equation}

In what follows we use the results in (\ref{DelEF}) -- (\ref{Gklrho}),
average over initial degenerate states $|lm \rangle$ and carry out some
angular algebra, more details of which are given in the Appendix.  We
introduce reduced matrix elements $\langle j'||\mathbf{X}(\lambda)||
j\rangle$, see (\ref{Wigner}) and then $\Delta E_1$ in (\ref{DelE1})
is given by
\begin{eqnarray}
\label{DelEF1}
\fl
    \Delta E_1 = \left[(2l+1)(2S+1)\right]^{-\frac{1}{2}}
    \langle S ||\mathbf{T}(\lambda)|| S \rangle
    \langle l ||\mathbf{U}(\lambda)|| l \rangle
    \int_0^\infty F_{kl\Lambda S}(r) V_{\rm{p}}(r) F_{kl\Lambda S}(r) \rmd r
    \nonumber \\  \hspace*{10em}
\end{eqnarray}
for $\lambda = 0$ and zero otherwise.  On using (\ref{FH0}), (\ref{Wigner})
and (\ref{C3j}), $\Delta E_2$ in (\ref{DelEF}) becomes
\begin{eqnarray}
\label{DelEF2}
\fl
    \Delta E_2 = \left[(2l+1)(2\lambda+1)(2S+1)\right]^{-1} \sum_{l'S'}
    |\langle S'||\mathbf{T(\lambda)}|| S \rangle|^2 \;
    |\langle l'||\mathbf{U(\lambda)}|| l \rangle|^2
    \nonumber \\ \lo \times
    \left[\int_0^\infty G_{kl'\Lambda'S'}(r) V_{\rm{p}}(r)
    F_{kl\Lambda S}(r) \rmd r \right.
    \nonumber \\ \lo
    - \left. \delta_{l'l}\, \delta_{S'S}
    \int_0^\infty F_{kl\Lambda S}(r) V_{\rm{p}}(r) F_{kl\Lambda S}(r) \rmd r
    \int_0^\infty G_{kl\Lambda S}(r) F_{kl\Lambda S}(r) \rmd r \right] \,.
\end{eqnarray}

\section{The spin-dipole interaction}

The interaction between the electronic-spin magnetic-dipole moments of each
atom produces the spin-dipole interaction Hamiltonian
\begin{equation}
\label{Hsd}
    H_{\rm{sd}} = V_{\rm{p}}(r)\,\frac{1}{\hbar^2}
    \left[3\,(\mathbf{S}_1\cdot\hat{\bi{r}})(\mathbf{S}_2
    \cdot\hat{\bi{r}})-\mathbf{S}_1\cdot\mathbf{S}_2\right] \,,
\end{equation}
where $\mathbf{S}_1$ and $\mathbf{S}_2$ are the electronic-spin operators for
the two atoms and $\hat{\bi{r}}$ is a unit vector directed along the
internuclear axis.  The function $V_{\rm{p}}(r)$ is defined by
\begin{equation}
\label{VpB}
    V_{\rm{p}}(r) = - \frac{\beta}{r^3}\,; \qquad \beta = \alpha^2 \left(
    \frac{\mu_{\rm{e}}}{\mu_{\rm{B}}}\right)^2 E_{\rm{h}} a_0^3 \,,
\end{equation}
where $\alpha$ is the fine structure constant, $a_0$ is the Bohr radius,
$(\mu_{\rm{e}}/\mu_{\rm{B}}) = 1.00115965$ is the electron magnetic moment and
$E_{\rm{h}}$ is the Hartree energy (= 1 a.u.).  The perturbation $H_{\rm{sd}}$
in (\ref{Hsd}) can easily be identified with $H_{\rm{p}}$ in (\ref{Hp}) since
\begin{equation}
\label{SS1S2}
    \mathbf{S}_1 \mathbf{S}_2 = \frac{1}{2}\, \left[\mathbf{S S} -
    \mathbf{S}_1\mathbf{S}_1 - \mathbf{S}_2\mathbf{S}_2 \right] \,,
\end{equation}
where $\mathbf{S} = \mathbf{S}_1 + \mathbf{S}_2$ is the operator for the
total spin.  Therefore in (\ref{Hp}), $\lambda = 2$ and
\begin{equation}
\label{TSS1S2}
    \mathbf{T}(2) \equiv \mathcal{S}(2)-\mathcal{S}_1(2)-\mathcal{S}_2(2) \,;
    \qquad
    \mathbf{U}(2) \equiv \frac{1}{2}(3\hat{\bi{r}}\hat{\bi{r}}-\mathbf{I}) \,,
\end{equation}
where $\mathbf{I}$ is the unit dyadic.  In (\ref{TSS1S2}), $\mathcal{S}(2)$,
$\mathcal{S}_1(2)$, $\mathcal{S}_2(2)$ and $\mathbf{U}(2)$ are irreducible
tensors with components
\begin{eqnarray}
\label{SS1S2q}
    \mathcal{S}(2\;q) =
    \frac{1}{\hbar^2}\left(\frac{4\pi}{5}\right)^{\frac{1}{2}}
    S^2 \;Y_{2\,q}(\hat{\mathbf{S}}) \,; \nonumber \\
    \mathcal{S}_i(2\;q) =
    \frac{1}{\hbar^2}\left(\frac{4\pi}{5}\right)^{\frac{1}{2}}
    S_i^2\; Y_{2\,q}(\hat{\mathbf{S}}_i) \,; \qquad i = 1,2
\end{eqnarray}
and
\begin{equation}
\label{U2q}
    U(2\;q) = \left(\frac{4\pi}{5}\right)^{\frac{1}{2}} Y_{2\,q}
    (\hat{\bi{r}}) \,.
\end{equation}
In (\ref{SS1S2q}) and (\ref{U2q}), functions of the type
$Y_{2\,q}(\hat{\mathbf{x}})$ are spherical harmonics and explicit
expressions for $\langle S'||\mathbf{T}(2)|| S \rangle$ and
$\langle l'||\mathbf{U}(2)|| l \rangle$ are given in (\ref{reducS}) --
(\ref{reduc}).

\section{Spin-polarized metastable helium atoms}

For the case of metastable helium atoms, $\Lambda = 0, S_i = 1; i = 1,2$ and so
the adiabatic potentials required for the ${}^1\Sigma^+_g$ and ${}^5\Sigma^+_g$
molecular states are the potentials $V_{00}(r)$ and $V_{02}(r)$.  Initially the
atoms are spin polarized so that $S = 2$ and $M_S = 2$ and then collisions take
place that produce final states with $S'= 2, M'_S = 0,\pm 1,\pm 2$ and $S'= 0,
M'_S = 0$.  For the ${}^5\Sigma^+_g$ state we use the analytical potential of
St\"arck and Meyer \cite{Starck1994} which has a scattering length of $157a_0$
and supports 15 bound states.  For the ${}^1\Sigma^+_g$ potential, we use the
results obtained by M\"uller \etal \cite{Muller1991} for $r < 12a_0$ and for $r
\geq 12a_0$, the potential is matched smoothly onto the long-range form
$V_{02}(r)-V_{\rm{exch}}(r)$ where $V_{\rm{exch}}(r)=A\exp(-\gamma r)$
\cite{Venturi1999,Leo2001}.  If the spin polarization is destroyed, there is a
high probability of Penning and associative ionization and subsequent loss of
atoms from the trap.  We model this loss by using a complex optical potential
of the form $V_{\Lambda'S'}(r) = V_{00}(r) - \case12\rmi \Gamma_{00}(r)$.  Two
forms for $\Gamma_{00}(r)$ are used; $\Gamma_{\rm{M}}(r)$ a least squares fit
to the tabulated results in \cite{Muller1991} and the simpler form
$\Gamma_{\rm{GMS}}(r)=0.3\exp(-r/1.086)$ of Garrison \etal \cite{Garrison1973}
which decreases more rapidly as $r$ increases and does not decrease for small
values of $r$.

For this case, the change in energy of the states with $l = 0$ is obtained
from (\ref{DelEF2}), (\ref{VpB}) and (\ref{special}), i.e.
\begin{equation}
\label{Delsd}
    \Delta E_2 = \frac{2}{5} \beta^2 \int_0^\infty F_{k002}(r) \frac{1}{r^3}
    \left[ G_{k200}(r) + 7\,G_{k202}(r) \right] \rmd r \,.
\end{equation}
In \cite{Peach2004}, it was shown that since the effective range of the
bound-state wave function is typically $10^3 a_0$ to $10^4 a_0$, the wave
function $F_{kl\Lambda S}$ for a trap state could be replaced by a free-wave
function of the same energy and an excellent value for the energy obtained.
Therefore in this application, the energy shifts and widths are calculated
using both bound-state ($\Delta_{\rm{trap}} = 1$) and free-wave solutions
($\Delta_{\rm{trap}} = 0$) of (\ref{Fkl}) for $F_{k002}$ to test further
the validity of the free-wave approximation.  Energy shifts and widths are
also calculated for trap states with $l = 2$.

\section{Numerical calculations}

The unperturbed eigenvalue equation in the form (\ref{Fkl}) or (\ref{Fklrho})
was solved using the two computational methods described in \cite{Peach2004}.
The first combines the use of quantum defect theory, numerical integration and
an iterative procedure (QDT) and in the second a direct numerical solution is
obtained using a discrete variable representation (DVR) of the kinetic energy
operator and a scaled radial coordinate grid.  The DVR method is easily
modified to solve (\ref{Gklrho}) for the perturbed functions $G_{kl\Lambda
S}(\rho)$.  A general real invertible transformation of the radial variable
$\rho$ given by
\begin{equation}
\label{rho}
    t=u(\rho); \qquad \rho = u^{-1}(t) \equiv U(t)
\end{equation}
is introduced so that (\ref{Gklrho}) becomes
\begin{equation}
\label{Gt}
    \left[-f^2\frac{\rmd^2}{\rmd t^2}f^2 + Q(t)\right]
    \tilde G(t) = \tilde W(t)\,\tilde F(t) \,,
\end{equation}
where
\begin{eqnarray}
\label{ft}
    f(t) \equiv \left[\frac{\rmd U}{\rmd t}\right]^{-1/2};
    \qquad \tilde F(t) \equiv \frac{F_{kl\Lambda S}[U(t)]}{f(t)} \,;
    \nonumber \\
    \tilde G(t) \equiv \frac{G_{kl'\Lambda'S'}[U(t)]}{f(t)} \,;
    \qquad \tilde W(t) \equiv \frac{W[U(t)]}{f(t)}
\end{eqnarray}
and
\begin{equation}
\label{Pt}
    Q(t) = \frac{l'(l'+1)}{\rho^2} + \rho^2 + \frac{2 V_{\lambda'S'}(\rho)}
    {\hbar\omega} + f^3 \frac{\rmd^2f}{\rmd t^2} - 4n^* \,.
\end{equation}
The DVR is constructed by using a finite set of basis functions $\{\phi_m(t)\}$
and coordinate points $\{t_m\}$ over the interval $[t_1,t_N]$, so that the
differential equation (\ref{Gt}) is transformed into the matrix eigenvalue
equation
\begin{equation}
\label{dvr}
    \sum_{j=1}^{N} \left[f^2(t_i)T_{ij}f^2(t_j) + Q(t_i)\,\delta_{ij}
    \right] \tilde G(t_j) = \tilde W(t_i)\,\tilde F(t_i) \,,
\end{equation}
where $i = 1,2,\ldots N$.  The matrix element $T_{ij}$ of the kinetic energy
operator $T = -\rmd^2/\rmd t^2$ obtained using a Fourier basis is given in
\cite{Peach2004} and we choose $\rho_1\xi = 2a_0$ and $\rho_N=15$.  The
scaling is given by
\begin{equation}
    t = u(\rho) = \left(\frac{\rho}{\zeta }\right)^{1/p}; \qquad
    \rho = U(t) = \zeta t^p
\end{equation}
and we choose $\zeta = 20$ and $p=10$ so that about 17\% of the scaled mesh
points lie between $\rho_1$ and $\zeta$ where the interatomic and spin-dipole
interactions are significant.  Four to five digit convergence is obtained for
the perturbed energies with $N=2000$.

Alternatively, having determined the values of $n^*$ for the bound states,
(\ref{Fkl}) and (\ref{Gkl}) are solved for the correct energies with
$\Delta_{\rm{trap}} = 0$.  The previous numerical procedure QDT can be
readily modified for this purpose and equations (\ref{Fkl}) and (\ref{Gkl})
are integrated numerically using the Numerov algorithm.  The solution of
(\ref{Gkl}) contains, in general, a particular integral plus a complementary
function which is some linear combination of $F^{\rm{R}}_{l'}(r)$ and
$F^{\rm{I}}_{l'}(r)$ as defined by (\ref{FRFI}) with $l\Lambda S$ replaced
by $l'\Lambda'S'$.  The method, to be labelled as QDTF, is as follows.
Equation (\ref{Fkl}) is integrated outwards from the origin to some
$r = r_{\rm{max}}$ and $F^{\rm{R}}_{l'}(r)$ and $\rmd F^{\rm{R}}_{l'}(r)/\rmd r$
are obtained.  On matching to solutions that are asymptotically plane waves
using methods similar to those described in \cite{Peach2004}, we obtain
$\delta_{l'}(k)$.  Then $F^{\rm{I}}_{l'}(r)$ and $\rmd F^{\rm{I}}_{l'}(r)/\rmd r$
are calculated at $r = r_{\rm{max}}$, c.f. (\ref{FRFI}), so that (\ref{Fkl})
can be integrated inwards to obtain the irregular solution.  Integration of
(\ref{Gkl}) requires some care in dealing with the multiples of
$F^{\rm{R}}_{l'}(r)$ and $F^{\rm{I}}_{l'}(r)$ that build up in the Numerov
integration in both the outward and inward directions.  In the integration of
(\ref{Gkl}) outwards, multiples of $F^{\rm{R}}_{l'}(r)$ are removed at each
integration step determined so that the function at the current point is zero
and this specifies $G_{\rm{out}}(r)$.  The function $G_{\rm{in}}(r)$ is obtained
by integrating inwards and a multiple of $F^{\rm{I}}_{l'}(r)$ is subtracted at
the end to make the function zero at the innermost point.  Finally, a multiple of
$F^{\rm{R}}_{l'}(r)$ is added to $G_{\rm{out}}(r)$ so that the solution matches
$G_{\rm{in}}(r)$ at $r \approx 6 a_0$ and this procedure is very insensitive
to the precise choice of matching point.  This completes the specification
of the solution $G_{k'l'\Lambda'S'}(r)$ of the inhomogenous equation (\ref{Gkl}).
It is found that for $r_{\rm{max}} = 1000 a_0$ convergence in the evaluation
of the radial integrals in (\ref{Delsd}) has been obtained correct to three
significant figures.

\section{Results and discussion}

The scaled energy shifts and widths $\Delta n^*$ for the five lowest trap
states with $l=0$ have been calculated using the DVR method for trapping
frequencies $\nu = \omega/2\pi$ ranging from 1 kHz to 10 MHz.  Calculations
have also been carried out for frequencies 100 kHz, 1 MHz and 10 MHz using the
QDTF approximation and the agreement with the DVR results is very satisfactory.
Differences for transitions to the $S'=0$ states range from 3\% for the lowest
state to 0.4\% for the highest, and at 10 MHz are always $\leq$ 1\%.  For
transitions to the $S'=2$ state the differences are even less, they are $\leq$
0.3\% except for the lowest state at 100 kHz where the difference is 2\%.  It
is clear why the QDTF approximation is not so good for the lowest states as
the range of $r$ over which the bound-state wave function can be approximated
closely by a free wave decreases with decreasing energy.  The QDTF 
approximation
has also been used in (\ref{DelEF2}) to investigate widths and shifts for $l=2$
and they prove to be negligible.  This is not surprising as the perturbation
weights the inner region strongly where the initial wave functions with $l \neq
0$ are very small.

The results shown in table 1 have been obtained using the DVR method.  The
contributions to $\Delta n^*$ of the terms with $S'=0,2$ in (\ref{Delsd}) are
also shown, together with the effective quantum number $n^*$ and the lifetime
$\tau$ given by
\begin{equation}
\label{tau}
    \tau =-1/(4\pi\nu\Im\Delta n^*) \,.
\end{equation}
The contributions from $S'= 0$ are sensitive to the form used for
$\Gamma_{00}(r)$.  In table 1 we also show lifetimes obtained using both
$\Gamma_{\rm{M}}(r)$ and $\Gamma_{\rm{GMS}}(r)$.  The energy shifts increase
with trapping frequency and $n^*$, but the fractional shifts $\Delta n^*/n^*$
decrease with $n^*$ and are of the order $10^{-5}$ to $10^{-7}$.  The energy
shift arises predominantly from the transitions to $S'=2$ states.  As expected,
the lifetimes $\tau_{\rm{GMS}}$ are shorter than the $\tau_{\rm{M}}$, as
$\Gamma_{\rm{GMS}}(r)$ is larger than $\Gamma_{\rm{M}}(r)$ for all $r \leq
5a_0$.  This dependence suggests that an experimental study of lifetimes as a
function of trap frequency could yield an improved knowledge of the potential
representing the decay channel.

The decay widths and lifetimes depend strongly on trapping frequency and the
lifetimes range from the order of $10^{4}$ s for 1 kHz traps to the order
of 15 ms for 10 MHz traps.
The lifetimes for the lower frequencies are greater than both the metastable
helium lifetime of 8000 s and the typical lifetimes of experimental
traps which are of the order of seconds \cite{Muller1997}.
Over the range of frequencies investigated the real and imaginary
parts of $\Delta n^*$ can be quite closely fitted to the analytic form
\begin{equation}
\label{Dnsfit}
    \Delta n^* = -A \sqrt{x}/(1+Bx+Cx^2+Dx^3) \,; \qquad x = \nu n^* \,,
\end{equation}
where $\nu$ is in units of MHz and
\begin{eqnarray}
\label{paramr}
    A =  4.0144 \times 10^{-6} \,; \qquad
    B =  1.8806 \times 10^{-2} \,; \nonumber \\
    C =  3.0740 \times 10^{-4} \,; \qquad
    D = -3.3204 \times 10^{-6}
\end{eqnarray}
for $\Re\Delta n^*$.  For $\Im\Delta n^*$
\begin{eqnarray}
\label{parami}
    A =  1.7294 \times 10^{-7} \,; \qquad
    B =  1.7145 \times 10^{-2} \,; \nonumber \\
    C =  3.3987 \times 10^{-4} \,; \qquad
    D = -3.7630 \times 10^{-6}
\end{eqnarray}
and the parametric fit to the lifetimes is then directly obtained from
(\ref{tau}), (\ref{Dnsfit}) and (\ref{parami}).

In conclusion, second-order perturbation theory with a rather general form for
the perturbation and valid for states with non-zero angular momentum has been
developed here.  These results are important and have been obtained so that
the analysis can be directly applied to other perturbations in future.

\appendix

\section*{Appendix}

\setcounter{section}{1}

The book on Angular Momentum by Edmonds \cite{Edmonds1974} is the basic
reference for the notation, equation numbers and tables quoted below.  The
Wigner-Eckhart theorem, see (5.4.1), defines the reduced matrix elements
$\langle j'||\mathbf{X}(\lambda)|| j \rangle$ using the relation
\begin{equation}
\label{Wigner}
    \langle j'm'_j|X(\lambda \;q)| j m_j \rangle = (-1)^{j'-m'_j}
    \langle j'||\mathbf{X}(\lambda)|| j \rangle
    \Threej{j'}{\lambda}{j}{-m'_j}{q}{m_j} \,,
\end{equation}
where {\footnotesize$\pmatrix{a & b & c\cr d & e & f\cr}$} is a $3-j$
symbol and the average over the degenerate initial states $|lm \rangle$
is carried out using
\begin{equation}
\label{C3j}
\fl
    (2j+1)^{-1} \sum_{m'_jm_j} \Threej{j'}{\lambda}{j}{-m'_j}{q'}{m_j}
    \Threej{j'}{\lambda}{j}{-m_j'}{q}{m_j} = [(2j+1)(2j'+1)]^{-1}
    \,\delta_{q'q} \,,
\end{equation}
see (3.7.8).  The states $|S M_S \rangle$ and $|S_i M_{S_i} \rangle$ are
eigenvectors of the operators $\mathbf{S}^2$, $S_z$, $\mathbf{S}_i^2$,
and $S_{iz}, i=1,2$.  Therefore using (\ref{Wigner}), (4.6.3) and
table 2, the reduced matrix elements required are given by
\begin{eqnarray}
\label{reducS}
\fl
    \langle S'_1 S'_2 S'||\mathcal{S}(2)|| S_1 S_2 S \rangle
    = \frac{1}{4}\left[(2S+3)(2S+2)(2S+1)(2S)(2S-1)\right]^{\frac{1}{2}}
    \delta_{S'S} \,;
    \nonumber \\
\fl
    \langle S'_i |\mathcal{S}_i(2)|| S_i \rangle = \frac{1}{4}
    \left[ (2S_i+3)(2S_i+2)(2S_i+1)(2S_i)(2S_i-1)\right]^{\frac{1}{2}}
    \delta_{S'_i S_i} \,; \qquad i=1,2
\end{eqnarray}
and
\begin{equation}
\label{reducl}
    |\langle l'||\mathbf{U}(2)|| l \rangle|^2 = (2l+1)(2l'+1)
    \left|\Threej{l'}{2}{l}{0}{0}{0}\right|^2 \,.
\end{equation}
Finally, using (7.1.7) and (7.1.8), the reduced matrix elements of
the spin operators $\mathcal{S}_1$ and $\mathcal{S}_2$ in the coupled
representation are given by
\begin{eqnarray}
\label{reduc}
\fl
    \langle S'_1 S'_2 S'||\mathcal{S}_1(2)|| S_1 S_2 S \rangle
    = (-1)^{S'_1+S_2+S} \left[(2S+1)(2S'+1)\right]^{\frac{1}{2}}
    \Sixj{S'_1}{S'}{S_2}{S}{S_1}{2}
    \nonumber \\ \lo \times
    \langle S'_1 ||\mathcal{S}_1(2)|| S_1 \rangle\, \delta_{S'_2 S_2} \,;
    \nonumber \\
\fl
    \langle S'_1 S'_2 S'||\mathcal{S}_2(2)|| S_1 S_2 S \rangle
    = (-1)^{S_1+S_2+S'} \left[(2S+1)(2S'+1)\right]^{\frac{1}{2}}
    \Sixj{S'_2}{S'}{S_1}{S}{S_2}{2}
    \nonumber \\ \lo \times
    \langle S'2 ||\mathcal{S}_2(2)|| S_2 \rangle\, \delta_{S'_1 S_1} \,,
\end{eqnarray}
where {\footnotesize$\left\{\matrix{a & b & c\cr d & e & f\cr}\right\}$}
is a $6-j$ symbol.

The following important special cases are easily derived from
(\ref{Wigner}) -- (\ref{reduc}) and tables 2 and 5\,:
\begin{equation}
\label{special}
\fl
    | \langle 0 ||\mathbf{T}(2)|| 2 \rangle |^2 = 10 \,; \qquad
    | \langle 2 ||\mathbf{T}(2)|| 2 \rangle |^2 = 70 \,; \qquad
    | \langle 2 ||\mathbf{U}(2)|| 0 \rangle |^2 = 1 \,.
\end{equation}

\Tables

\begin{table}
\caption{Effective quantum numbers $n^* = E/(2\hbar \omega)$, shifts
$\Delta n^*$ (in units of $10^{-5}$) and state lifetimes $\tau_{\rm{M}}$
for harmonic traps with frequencies ranging from 1 kHz to 10 MHz.  Also
shown are the lifetimes $\tau_{\rm{GMS}}$ for comparison.  Numbers in
parentheses denote powers of 10.}

\begin{tabular}{@{}lllllll}
\br
    $n'_r$ & \centre{1}{$n^*$} & \centre{1}{$\Delta n^*(S'=0)$} &
    \centre{1}{$\Delta n^*(S'=2)$} & \centre{1}{$\Delta n^*(\rm{Total})$} &
    \centre{1}{$\tau_{\rm{M}}$(s)} & \centre{1}{$\tau_{\rm{GMS}}$(s)} \\
\mr
\centre{6}{$\nu$ = 1 kHz} \\
\mr
     0 &  0.7521 & $-0.0007-0.0005\rmi$ & $-0.0105$ & $-0.0112-0.0005\rmi$
     & 1.65(4) & 1.42(4) \\
     1 &  1.7531 & $-0.0010-0.0007\rmi$ & $-0.0158$ & $-0.0168-0.0007\rmi$
     & 1.10(4) &  9.47(3) \\
     2 &  2.7539 & $-0.0013-0.0009\rmi$ & $-0.0197$ & $-0.0210-0.0009\rmi$
     & 8.83(3) &  7.58(3) \\
     3 &  3.7546 & $-0.0015-0.0011\rmi$ & $-0.0229$ & $-0.0244-0.0011\rmi$
     & 7.57(3) &  6.50(3) \\
     4 &  4.7551 & $-0.0017-0.0012\rmi$ & $-0.0258$ & $-0.0275-0.0012\rmi$
     & 6.73(3) &  5.78(3) \\
\mr

\centre{6}{$\nu$ = 10 kHz} \\
\mr
     0 &  0.7566 & $-0.0022-0.0015\rmi$ & $-0.0335$ & $-0.0356-0.0015\rmi$
     & 5.20(2) & 4.47(2) \\
     1 &  1.7599 & $-0.0032-0.0023\rmi$ & $-0.0500$ & $-0.0532-0.0023\rmi$
     & 3.48(2)  & 2.99(2) \\
     2 &  2.7624 & $-0.0040-0.0029\rmi$ & $-0.0623$ & $-0.0663-0.0029\rmi$
     & 2.79(2) & 2.39(2) \\
     3 &  3.7644 & $-0.0047-0.0033\rmi$ & $-0.0726$ & $-0.0773-0.0033\rmi$
     & 2.39(2) & 2.05(2) \\
     4 &  4.7662 & $-0.0053-0.0037\rmi$ & $-0.0816$ & $-0.0869-0.0037\rmi$
     & 2.13(2) & 1.83(2) \\
\mr

\centre{6}{$\nu$ = 100 kHz} \\
\mr
     0 &  0.7711 & $-0.0069-0.0049\rmi$ & $-0.1073$ & $-0.1142-0.0049\rmi$
     & 1.62(1) & 1.39(1) \\
     1 &  1.7815 & $-0.0102-0.0073\rmi$ & $-0.1588$ & $-0.1691-0.0073\rmi$
     & 1.09(1) & 9.37   \\
     2 &  2.7893 & $-0.0127-0.0091\rmi$ & $-0.1971$ & $-0.2098-0.0091\rmi$
     & 8.78 & 7.54   \\
     3 &  3.7957 & $-0.0147-0.0105\rmi$ & $-0.2288$ & $-0.2435-0.0105\rmi$
     & 7.56 & 6.50 \\
     4 &  4.8014 & $-0.0165-0.0118\rmi$ & $-0.2562$ & $-0.2727-0.0118\rmi$
     & 6.75 & 5.80 \\
\mr

\centre{6}{$\nu$ = 1 MHz} \\
\mr
     0 &  0.8181 & $-0.0225-0.0160\rmi$ & $-0.3493$ & $-0.3718-0.0160\rmi$
     & $4.96(-1)$  & $4.26(-1)$ \\
     1 &  1.8502 & $-0.0318-0.0229\rmi$ & $-0.4960$ & $-0.5277-0.0229\rmi$
     & $3.48(-1)$  & $2.99(-1)$ \\
     2 &  2.8738 & $-0.0382-0.0275\rmi$ & $-0.5967$ & $-0.6349-0.0275\rmi$
     & $2.89(-1)$  & $2.48(-1)$ \\
     3 &  3.8930 & $-0.0430-0.0311\rmi$ & $-0.6740$ & $-0.7170-0.0311\rmi$
     & $2.56(-1)$ & $2.19(-1)$ \\
     4 &  4.9096 & $-0.0469-0.0340\rmi$ & $-0.7362$ & $-0..7831-0.0340\rmi$
     & $2.34(-1)$ &  $2.01(-1)$ \\
\mr
\centre{6}{$\nu$ = 10 MHz} \\
\mr
     0 &  0.9703 & $-0.0645-0.0469\rmi$ & $-1.0145$ & $-1.0791-0.0469\rmi$
     & $1.70(-2)$  & $1.46(-2)$ \\
     1 &  2.0509 & $-0.0720-0.0533\rmi$ & $-1.1468$ & $-1.2188-0.0533\rmi$
     & $1.49(-2)$  & $1.28(-2)$   \\
     2 &  3.1040 & $-0.0732-0.0550\rmi$ & $-1.1783$ & $-1.2516-0.0550\rmi$
     & $1.45(-2)$  & $1.24(-2)$  \\
     3 &  4.1442 & $-0.0725-0.0551\rmi$ & $-1.1777$ & $-1.2502-0.0551\rmi$
     & $1.44(-2)$  & $1.24(-2)$ \\
     4 &  5.1767 & $-0.0710-0.0546\rmi$ & $-1.1648$ & $-1.2358-0.0546\rmi$
     & $1.46(-2)$  & $1.25(-2)$ \\
\br
\end{tabular}
\end{table}
\end{document}